\documentclass[parskip=full, 10pt]{scrartcl}
\usepackage[left=2.2cm,right=2.2cm,
    top=2cm,bottom=2cm,bindingoffset=0cm]{geometry}
\usepackage{tipa}

\usepackage[numbers,sort&compress]{natbib}
\usepackage{hyperref}
\usepackage{booktabs}
\usepackage{mdframed}
\hypersetup{
    colorlinks=true,
    urlcolor=blue,
    citecolor=blue,
    linkcolor=blue
}
\usepackage{lipsum}
\usepackage{array}
\usepackage{wrapfig}
\usepackage{natbib}
\usepackage{enumerate}
\usepackage{authblk}

\usepackage{forest}
\usepackage{soul}
\DeclareUnicodeCharacter{2212}{-}
\usepackage{multirow}
\usepackage{graphicx}
\usepackage{dcolumn}
\usepackage{bm}
\usepackage{soul}
\usepackage{forest}
\usepackage{booktabs}
\usepackage{amsmath}
\usepackage{amssymb}
\usepackage{subfigure}
\usepackage{tikz}
\tikzset{
    ncbar angle/.initial=90,
    ncbar/.style={
        to path=(\tikztostart)
        -- ($(\tikztostart)!#1!\pgfkeysvalueof{/tikz/ncbar angle}:(\tikztotarget)$)
        -- ($(\tikztotarget)!($(\tikztostart)!#1!\pgfkeysvalueof{/tikz/ncbar angle}:(\tikztotarget)$)!\pgfkeysvalueof{/tikz/ncbar angle}:(\tikztostart)$)
        -- (\tikztotarget)
    },
    ncbar/.default=0.2cm,
}
\tikzset{square left brace/.style={ncbar=0.1cm}}
\tikzset{square right brace/.style={ncbar=-0.1cm}}

\usepackage{subcaption}
\newcommand{\dd}{\mathop{}\mathopen{}\mathrm{d}}

\makeatother
\title{Inferring Interpretable Models of Fragmentation Functions using Symbolic Regression}

\author{\large{Nour Makke}}
\author{Sanjay Chawla}

\affil{Qatar Computing Research Institute, HBKU, Doha}

\date{\large{January 12, 2025}}

\begin{document}
\maketitle


\begin{abstract}
\begin{center}
\textbf{\abstractname} \\[2ex]
\end{center}
Machine learning is rapidly making its path into natural sciences, including high-energy physics. We present the first study that infers, directly from experimental data, a functional form of fragmentation functions. The latter represent a key ingredient to describe physical observables measured in high-energy physics processes that involve hadron production, and predict their values at different energy. Fragmentation functions can not be calculated in theory and have to be determined instead from data.  Traditional approaches rely on global fits of experimental data using a pre-assumed functional form inspired from phenomenological models to learn its parameters. This novel approach uses a ML technique, namely symbolic regression, to learn an analytical model from measured charged hadron multiplicities. The function learned by symbolic regression resembles the Lund string function and describes the data well, thus representing a potential candidate for use in global FFs fits. This study represents an approach to follow in such QCD-related phenomenology studies and more generally in physics. 
\end{abstract}

\vspace{-.4cm}
\section{Introduction}
\vspace{-.4cm}

Fragmentation functions represent a key ingredient in the description of hadron production cross sections in various high-energy physics (HEP) processes, i.e., lepton-nucleon, nucleon-nucleon, and nuclei-nuclei collisions. They provide a quantitative description of the hadronization mechanism~\cite{webber1994hadronization}, which is intrinsically non-perturbative in the Quantum Chromodynamics (QCD) theory. FFs are not calculable in perturbative QCD, and their determination fully relies on physical observables measured in high-energy physics experiments, e.g., the large hadron collider at CERN~\cite{ALICE:2008ngc,Schmitt:1998rq,CMS:2008xjf,ATLAS:2008xda}, among others. The current methodology relies on global QCD fits~\cite{PhysRevD.75.114010,PhysRevD.105.L031502}, commonly referred to as ``FFs parameterizations'', where a pre-assumed functional form of FFs is fit to a wide range of physical observables to learn its parameters by involving the DGLAP evolution equations~\cite{ALTARELLI1977298} which considers the different energy scales of the experimental measurements. 
FFs represent a key ingredient to describe hadron production in all HEP processes at the running experiments at the large hadron collider (LHC) at CERN, and to make predictions for the next generation of experiments such as the future Electron Ion Collider (EIC) at the Berkeley National Laboratory (BNL) and the future Circular Colider (FCC) at CERN which will run at significantly higher energies (with center-of-mass energy of 100 TeV versus 14 TeV at LHC) thus covering new regions of the kinematic phase space.

It is mandatory to question, in the fast-evolving AI era, whether ML could assist in inferring a functional form of FFs directly from data rather than pre-assuming a function, and, most importantly, if the function learned using AI tools is interpretable, human-understandable, and how it compares to designated functions. 
This paper reports the first study to infer a functional form of fragmentation functions from experimental data using an ML-based technique, i.e., symbolic regression (SR). 
The latter was introduced back in 1970 as an AI tool to automate scientific discovery using heuristics and succeeded in learning various fundamental physics laws (Ohm's law, Galileo's law, Ideal gas's law, etc.). Approaches to solving SR have evolved significantly from traditional search-based approaches (e.g., heuristic search and evolutionary algorithms) to modern learning-based approaches (e.g., ML-based methods) and hybrid techniques, as reviewed in~\cite{Makke2024,makkekdd}.
SR is recently showing remarkable efficiency in learning analytical models directly from data and is proving a potential candidate for an automated scientific discovery tool. 
Mathematical equations are interpretable and thus transparent models, in contrast to deep neural networks known as black-box models due to their full opacity.
Symbolic regression is particularly suited for physics applications since physical laws are expressed by mathematical equations. Its application to experimental data, however, is very limited~\cite{cranmersolar,Reinbold2021}, in particular in high-energy physics~\cite{10.1093/pnasnexus/pgae467,Tsoi_2024}, and was generally deployed on synthetic datasets in the majority of SR applications. Fragmentation functions represent an interesting and challenging application of symbolic regression because they can not be calculated in theory and are only determined by QCD fits, which adopt similar functional forms. This study serves two purposes: first, it tests the credibility of SR as a scientific discovery tool in physical science applications using experimental data which is naturally noisy; second, it compares what the data reveals about FFs with established functional forms.

Fragmentation functions~\cite{Metz_2016} represent, from a phenomenological point of view, the probability of a particular parton $i$ to transform into a charged hadron $h$ carrying a fractional energy $z$. In global QCD analyses~\cite{PhysRevD.75.114010,PhysRevD.105.L031502}, an FF is parameterized at an initial scale $Q_0$, cf. Eq.~\ref{eq:FF}, where the free parameters $N_i, \alpha_i, \gamma_j, \beta_{ij}$ are determined by the fit.
\begin{equation}\label{eq:FF}
    D_i^{h}(z,Q_0) = N_{i} z^{\alpha_i}\sum_{j=1}^{3}\gamma_{j}(1-z)^{\beta_{ij}}
\end{equation}
The component $(1-z)^{\beta}$ constraints the FFs at $z=1$, such that $D_i^h(z=1, Q_0) = 0$. This functional form is inspired by the Lund symmetric fragmentation function~\cite{ANDERSSON198331} given by:
\begin{equation}\label{eq:LundFF}
    f(z) \propto (1/z)(1-z)^{\alpha}\exp(-\beta m_{h}^2/z)
\end{equation}
Where $\alpha$ and $\beta$ are parameters that should be tuned to reproduce the experimental data.  It is worth noting that this study does not replace or eliminate the need for global QCD fits; it rather compliments them by suggesting a functional form of FFs that originates from data. 
Although SR is not a new technique, and despite significant advancements in SR techniques in recent years, successfully achieving this objective remains a challenging and uncertain endeavor.


\vspace{-.4cm}
\section{{Related Work}}
\vspace{-.4cm}

Fragmentation functions, and similarly parton distribution functions, were determined in previous studies~\cite{AbdulKhalek:2022laj,Bertone:2024taw,NNPDF:2021njg} that make use of ML, where a functional form of FFs is replaced by a deep neural network (DNN). DNNs have shown remarkable success~\cite{lecun95,Sherstinsky_2020,mnih2013playing,DBLP:journals/corr/VaswaniSPUJGKP17} in finding patterns in large and complex datasets and thus learning highly predictive models, leading to the deep learning (DL) revolution in early 2010 and allowing ML to make its pathway to nearly all domains, including natural sciences. DNNs are universal function approximations defined, for a given number of layers ($L$) with $L>1$, by the equation:
\begin{equation}\label{eq:dnn}
    f^{\mathrm{NN}}(z) = g(W^{[L]}z^{[L]} + b^{[L]}), \quad z^{[L]}=f^{\mathrm{NN}}(z^{[L-1]})~\text{and}~z^{[0]}=\mbox{x}
\end{equation}
Where $\mbox{x}\in\mathbb{R}^{d}$ is the input vector of features (or variables) of size $d$, $g$ is an activation function, and $(W,b)$ are sets of parameters whose numerical values are learned to best describe the ``training'' data. The function in Eq.~\ref{eq:dnn} becomes progressively opaque for increasing values of $L$\footnote{An extreme example is the OpenAI ChatGPT-4~\cite{chatgpt4}, for which have around $1.8$ Trillion parameters.}, making it impossible to reason about the relationship between the input $\mbox{x}$ and the prediction $y=f^{\mathrm{NN}}(\mbox{x})$, the reason why DNNs are called ``black-box'' models. 
Despite the remarkable predictive potential of DNNs, they are less suitable for use in physical sciences whose primary focus is to understand the universe through naturally occurring mechanisms and to transfer and unify knowledge, a goal that may be hard to achieve by learning large numerical models, also given the theory-experiment dual nature of physics. An in-depth discussion on the role of ML in sciences can be found in~\cite{hogg2024machinelearninggoodbad,makke_ml4ps}, and a review on the application of ML in physical sciences can be found in~\cite{RevModPhys.91.045002}.
An alternative approach to numerical models is to learn human-understandable, thus interpretable, models using the symbolic regression technique (cf. Sect.~\ref{sec:sr}). 
In contrast to previous ML-based studies of FFs where a black-box model was used to predict the numerical values of FFs, the present study uses black-box models to learn symbolic equations and, thus, functional forms of FFs which can be used in traditional global QCD fits.
A key advantage here is that no constraints or assumptions are made either at the model training or inference levels. 
In previous studies of FFs, training datasets are generated using phenomenological models of underlying physical mechanisms that make assumptions, whereas training data in this study consists of a large set of mathematical equations randomly generated using basic arithmetic and mathematical operations. Therefore, there is no assumption on the ``unknown'' underlying physical mechanism. In addition, the featured parameters in the inferred equation could have physical interpretation, thus allowing for a deeper understanding of the hadron formation mechanism and not just allowing us to make accurate numerical predictions of FFs.

\vspace{-.4cm}
\section{{Symbolic Regression and the Physical Sciences}}
\vspace{-.4cm}

The concept of SR, i.e., using data ``to discover an equation'' rather than ``fit the equation'', is as old as J.~Kepler's discovery, the 16th century, of the power law that describes the planetary motion. Modern SR was initiated by D.~Gerwin and P.~Langley in the 1970s and has been explored for decades in the context of scientific discovery by first developing Bacon~\cite{bacon1, 10.5555/29379}, which successfully uncovered fundamental physics laws from empirical data (Ohm's law, Galileo's law, etc.), followed by Coper~\cite{coper}, Fahrenheit~\cite{fahrenheit1,fahrenheit2}, and Lagrange~\cite{lagrange} developed to discover laws that govern the behavior of dynamical systems based on ideas from inductive logic programming. 
The rise of genetic algorithms and the newly introduced representation of mathematical equations using expression trees (cf. Sec.~\ref{sec:sr}) in the pioneering work by J.~Koza~\cite{koza:1994} has revived SR, which was since developed within the genetic programming community for a few decades. Although SR has been mainly developed for discovering physical laws, it has not been adopted in the physical sciences. This could be explained by the limitations encountered in both areas. For example, the accuracy reached in the current experimental measurement is in no way comparable to the accuracy that was reachable a few decades ago, and the efficiency of SR in processing high-dimensional datasets was limited due to computation resources in light of the exponentially growing size of the search space. Following the DL revolution, SR is re-emerging as a potential candidate to overcome the interpretability issue of black-box models and automate discovery in sciences.

The scientific approach focuses on learning about the latent structure of the world by developing fundamental physical theories. In physical sciences, there exist, in addition to fundamental theories, phenomenological models that describe some physical phenomena that can not be described by theory. SR could be used to learn either a fundamental law, part of a more general theory, or a phenomenological model. In both cases, however, the goal is to gain insights into the studied phenomena and not simply to fit data. 
This paper rigorously investigates the capability of SR to infer a functional form of FFs from experimentally measured (noisy) data and compares its structure and performance with established ones.
Although SR is a relatively old technique, and despite significant advancements in SR methods in recent years, successfully achieving this objective remains a challenging and uncertain endeavor. 

\vspace{-.4cm}
\section{Dataset and Method}\label{sec:sr}
\vspace{-.4cm}

The dataset comprises differential multiplicities of charged hadrons~\cite{COMPASS:2016xvm,COMPASS:2016crr} of different species (unidentified $h^{\pm}$, pions $\pi^{\pm}$, kaons $K^{\pm}$) measured in semi-inclusive deep inelastic scattering at the COMPASS experiment~\cite{COMPASS:2007rjf} at CERN. Data were collected by scattering a beam of muons of 160 (GeV/$c$) off a stationary deuterium target. Multiplicities are measured as a function of the hadron's fractional energy $z$ (i.e., $0.2<z<0.8$) in bins of the Bjorken scaling variable in the range $0.004<x<0.18$ and the virtual photon transfer momentum in the range $0.1 < y < 0.7$, as reported in~\cite{COMPASS:2016xvm,COMPASS:2016crr}. This dataset plays an instrumental role in constraining FFs in global QCD fits~\cite{PhysRevD.109.052004} thanks to its richness, where multiplicities are presented in a very detailed binning in the relevant kinematical variables. From a technical point of view, it encompasses multiple subsets that reveal a consistent fundamental structure while spanning diverse regions in the phase space. This mirrors multiple instances of SR to the same physics problem but with distinct data points. In addition, the effectiveness of the results can be easily verified for generalization within the same dataset and extended to other datasets, given the universality of FFs.

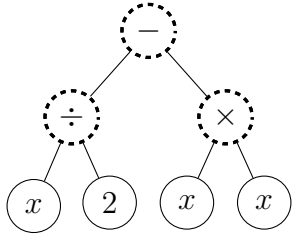
\begin{figure}[t]%
\centering
\begin{forest}
for tree = {circle, draw,
        minimum size=2.em,
        inner sep=0.5pt,
        font=\large,
        l sep=4mm,s sep=3mm
        },
        rbcirc 2/.style={%
            circle,
            fill=white,
            draw=black,
            dash pattern= on 2pt off 2pt,
            very thick,
            postaction={%
                draw=red,
                dash pattern= on 2pt off 2pt,
                dash phase=2pt,
                very thick,
            },
  },
[$-$,rbcirc 2[$\div$,rbcirc 2[$x$][2]][$\times$,rbcirc 2[$x$][$x$]]]
\end{forest}%
\caption{Expression-tree structure of the equation $f(x)=x/2-x^2$. Internal (and root) nodes are dashed lines and terminal nodes are full solid lines. Edge lines connect between operators and their respective sibling(s).}
\label{fig:treestructure}
\end{figure}

Symbolic regression (SR) aims to simultaneously learn models' structure $(f_{\theta})$ and parameters ($\theta$) directly from data, in contrast to DNNs where only models' parameters are learned. A mathematical equation is regarded as a unary-binary tree~\cite{10.5555/1623755.1623877} of symbols, whose nodes are operations and leaves are operands, as illustrated in the example of Fig.~\ref{fig:treestructure}. This representation allows expressing any equation as a sequence of symbols by traversing its tree, referred to as the Polish notation~\cite{polishnotation}. The latter is a mathematical notation in which operations precede operands, e.g., $f(z)=z^{\alpha}(1-z)^{\beta}\equiv \{*,\text{pow},z,\alpha,\text{pow},-,1,z,\beta\}$. A key distinction between SR and standard linear regression problems is in the discrete nature of the search space. The optimization problem in SR is defined over a discrete space of mathematical expressions, composed from a user-defined set of allowable mathematical operations, commonly referred to as the ``library'', e.g., $\mathcal{L} = \{\mathrm{add,~sub,~mul},~\mathrm{etc}.\}$; in general, SR has been shown to be an ``NP-hard" problem~\cite{virgolin2022symbolic}. SR methods have evolved significantly from traditional search-based approaches (e.g., heuristic search and evolutionary algorithms) to modern learning-based (e.g., transformer-based models) and hybrid techniques, as reviewed in~\cite{Makke2024,livingreview,makkekdd}.

We specifically choose the \textit{NeSymReS}~\cite{nesymres} method based on an encoder-decoder transformer architecture~\cite{srtnn}. Transformers were developed in natural language processing (NLP) to learn the context in text data by introducing attention blocks into an NN's architecture. The outstanding performance of transformers has quickly expanded their use beyond NLP to general sequential data, including time-series data. 
In the context of SR, transformers are configured as set-to-sequence models, taking as input a set of numerical data points and outputting a sequence of mathematical symbols.
Consider an univariate function $y(x)=x^{\alpha}(1-x)^{\beta}$ and a data set $\mathcal{D}=\{x,y\}_{i=1}^{n}$. In the training phase,
the numerical data is converted into a 16-bit binary representation and then passed to the encoder as a set of (2,16) matrices consisting of $n$ elements, where ``2'' denotes the number of variables, i.e., $x$ and $y$. The encoder applies a set transformer and multi-head pooling operations to output a high-dimensional latent embedding vector ($\mbox{z}$) of the set, and the actual size is determined by hyperparameters tuning to achieve the best performance during training. The ground-truth equation is converted into its skeleton form where all constants are replaced by placeholders, i.e., $e=x^{\circ}(1-x)^{\circ}$, and represented in a sequential prefix form with positional encoding following the standard decoder architecture introduced in~\cite{srtnn}, i.e., $\{*,\text{pow},x,\circ,\text{pow},-,1,x,\circ\}$. The decoder is then fed the skeleton $(e)$ with the positional embedding and the latent representation $(\mbox{z})$, and outputs a probability distribution over all the valid tokens, $P(e_{k+1}|e_{1:k},\mbox{z})$, where $(k+1)$ denotes the token to be predicted and ($1:k$) denotes the previously predicted tokens. The loss function is the standard cross-entropy loss. Its value is backpropagated through both the decoder and the encoder blocks, and their weights are updated. 
This process repeats across all training examples. The model is trained to reduce the average loss between the skeletons of the predicted equation and the ground-truth one. 

During inference, a new data set ($X,y$) is encoded into $\mbox{z}$, which is then passed through the decoder to create a sequence of symbols in an auto-regressive manner, i.e., each symbol generated is then appended to the input, and the next symbol is generated based on the new context. Finally, the generated skeleton equation is converted into a ``proper'' equation by replacing the constant tokens (``$\circ$'') with their numeric counterparts using non-linear optimization. Whereas the loss is defined between equations' skeletons in pre-training, it is minimized in the optimization of the numerical constants in inference mode.
Two pre-trained NeSymReS models are available. They can be directly used for inference without requiring the model to be trained from scratch for each new problem. This study uses the model pre-trained on 100 million datasets ($X,y,e$). Its parameters are loaded into the model, which is then called for in every inference problem.
The choice of a transformer-based SR here is mainly driven by the fact that learning the context in data holds significant meaning in physics, particularly in light of the causal nature of physical phenomena, where capturing correlations among variables is crucial. 

The complexity of the symbolic expressions (i.e., length of equations' sequences) represents an important challenge in SR due to possible overfitting. 
Foremost, any equation can generally be expressed in infinite ways (e.g., $\exp(-\alpha/x)$ is equivalent to $\exp(-\alpha/x +1−1)$ or $\exp(-1*(\alpha/x))$ or $(2−1)*\exp(-\alpha/x)$, etc.) and thus could be more or less complex.
The regularization of models' complexity can generally be enforced either in the definition of the training data, for example, by using simplest possible expressions in training,  or by adding a regularizer term to the loss function that penalizes long expression trees. The regularization task in this study is controlled by using training examples, where equations are succinct and do not include any additional terms that cancel out. Moreover, 
the complexity level, i.e., the length of the training equations, is handled in the training data generation phase~\cite{nesymres}, where a randomly generated expression tree has five or fewer non-leaf nodes, and longer expressions tend to be simplified into shorter ones such that the training dataset predominantly includes shorter expressions. As a result, the pretrained model is biased toward learning succinct, thus shorter expressions, preventing the addition of unnecessary parameters. A principled way of including regularization in SR is an open area
of research as SR is primarily used for automated scientific discovery and excessively penalization for short expression could hinder the discovery process.

\vspace{-.4cm}
\section{Analysis and Results}
\vspace{-.4cm}

We perform the analysis using differential multiplicities~\cite{COMPASS:2016xvm,COMPASS:2016crr} of positively and negatively charged hadrons ($h^{\pm}$), pions (${\pi^{\pm}}$), and kaons (${K^{\pm}}$) measured in nine $x$ bins and up to five $y$ bins within each $x$-bin, resulting in a total of 38 ($x,y$) bins. The limits of the three-dimensional binning are summarized in Tab.~\ref{tab:kinematics}. An SR problem takes as input $\mathcal{D}=\{\mbox{x},M^{h}\}_{i=1}^{n}$, where $\mbox{x}$ is a $d$-dimensional vector of kinematic variables with utmost $d=3$, and $n$ the size of $M^h$, and outputs an analytical equation $f(\mbox{x})$ that best describes $M^{h}(\mathrm{x})$. The analysis is performed using i) one-dimensional ($d=1$) multiplicities in Sec.~\ref{sec:1d} (i.e., $\mathcal{D}=\{z,M^{h}\}$) and Sec.~\ref{sec:1dff} (i.e., $\mathcal{D}=\{z,D^{h}\}$) where $D^h$ denotes extracted FFs, and ii) two-dimensional multiplicities ($d=2$) in Sec.~\ref{sec:2d} (i.e., $\mathcal{D}=\{z,x,M^{h}\}$), and results are presented and discussed within each section.

\vspace{-.4cm}
\subsection{{Learning univariate function from multiplicity values}}\label{sec:1d}
\vspace{-.25cm}

\begin{figure}[t]%
\centering
\includegraphics[width=18cm,height=12cm]{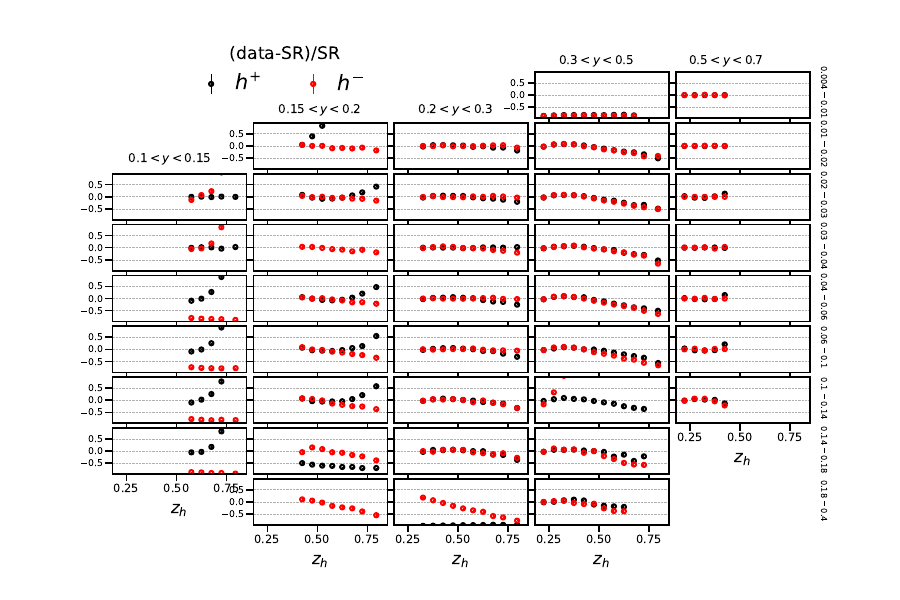} 
\caption{"(Data-SR)/SR" comparison for $h^{+}$ and $h^{-}$  multiplicity values from~\cite{COMPASS:2016xvm}. SR here refers to equations independently learned in individual ($x,y$) bins. The bins where a set of points is missing refer to cases for which normalization factors are missing, e.g., $f^{h^{+}}_{\text{SR}}=(\cos(2z))^3/z$ for $0.03<x<0.04$ and $f^{h^{+}}_{\text{SR}}=(\cos(2z))^5/z$ for $0.18<x<0.4$ for $0.15<y<0.2$, $f^{h^{+}}_{\text{SR}}=\exp(-z)/z^2$ for $0.18<x<0.4$ and $0.2<y<0.3$, and $f^{h^{-}}_{\text{SR}}=\exp(-az)/z^2$ for $0.1<x<0.14$ and $0.3<y<0.5$.} 
\label{fig:data_sr_hpm}
\end{figure}

Table~\ref{tab:results_hpm} summarizes the functions inferred independently from $h^{+}$ and $h^{-}$ multiplicities by applying SR (\textit{NeSymReS}) in individual kinematic bins $(x,y)$, resulting in 38 candidate functions for each charged hadron dataset. $n$ denotes the size of the input vector $M^{h^{+}/h^{-}}(z)$ per ($x, y$) bin, and the different values of $n$ in the range $0.3<y<0.5$ correspond to different $x$ bins. Various functional forms are repetitively learned across the $y$ bins, and they are all reported for completeness. 
Figure~\ref{fig:data_sr_hpm} illustrates the performance of the functions inferred by SR in terms of a “(data-model)/model” comparison as a function of $z$ for $h^{+}$ and $h^{-}$, where data denotes measured $M^{h}(z)$ and model denotes predicted $M^{h}(z)$ produced using the functions learned in individual ($x,y$) bins. The comparison of Fig.~\ref{fig:data_sr_hpm} visualizes the effectiveness of the functions learned by SR in describing the datasets used for their inference and mainly serves to explain our selection of the target function in terms of data description. The advantage of the model learned by SR with respect to established functional forms~\cite{PhysRevD.75.114010,PhysRevD.105.L031502} is further discussed at a later stage in the paper.
The kinematic bins where a set of points is partially shown or missing refer to cases for which normalization factors are missing in the learned models, i.e., $a=1$. 
The top performing functions that significantly describe the $z$ dependence of $M^{h^{+}/h^{-}}(z)$ and quantitatively match the data, and are associated with lower error are learned in the third $y$ bin, i.e., $0.2 < y <0.3$ ($3^{\text{rd}}$ column of Fig.~\ref{fig:data_sr_hpm}), with loss values of the order of $10^{-5}$ and up to $10^{-3}$, except for the last $x$ bin where $a=1$ for $h^{+}$, i.e., $f^{h^{+}}_{\text{SR}}=\exp(-z)/z^2$, and a trigonometric function is learned for $h^{-}$ with $a=1$, i.e., $f^{h^{-}}_{\text{SR}}=(z^2\tan(z+b))^{-1}$. 
It is worthy of note that the lowest error of $2.6\times 10^{-6}$ is obtained for the function $f^{h^{+}}(z;y_4)=a\left(1-b\cos(3z)\right)^{c}$ inferred by SR in the range $y\in[0.5,0.7]$ and five $x$ bins within $x\in[0.004,0.06]$. However, this function is not selected as top-performing mainly because it includes a cosine operation while data does not exhibit periodic behavior, and the $z$ range covered in the corresponding $y$ bin (i.e., $y_4$) is limited to $z<0.5$. This highlights the importance of the human-in-the-loop component in physics applications and generally sciences-related applications. Therefore, the top-performing functions are:
\begin{equation}\label{eq:top-performing}
\begin{split}
    & f_1(z) = a\exp(-bz)/z^2 \\
    & f_2(z) = a\exp(-bz)/(z-c)
\end{split}
\end{equation}
\begin{table}
    \centering
    \caption{Results of mathematical expressions inferred by SR (\textit{NeSymReS}) in individual kinematic ($x,y$) bins (38 bins in total) using $\mathcal{D}=\{z,M^{h}\}_{i=1}^{n}$ and presented in different $y$ ranges. ``NOF'' denotes the number of findings.}
    \begin{tabular}{p{2.5cm}p{1cm}lrr}
    \toprule
        $y$-range & $n$ & Expression & NOF($h^{+}$) & NOF($h^{-}$) \\
        \midrule
        \multirow{2}{*}{$0.1 < y < 0.15$} & \multirow{2}{*}{5} & $a\cos(bz)/z$ & 4 & 4 \\
        & &  $a\cos(bz)/(z+c)$ & 2 & - \\
        & & $a\cos(z)^b/z$ & - & 3 \\
        \midrule
        \multirow{2}{*}{$0.15 < y < 0.2$} & \multirow{2}{*}{8} & $(\cos(2z))^n/z$ & 4 & 1\\
        & & $\sin((b(cz-1)^2))^4$ & 4 & -\\
        & & $a\cos(z)^2/z^2$ & - & 7 \\        
        \midrule
        \multirow{2}{*}{$0.2 < y < 0.3$} & \multirow{2}{*}{10} & $a\exp(-bz)/z^2$ & 7 & 3\\
        & & $a\exp(-bz)/(z-c)$ & 1 & 4\\
        & & $1/(z^2\tan(z)+b)$ & - & 1 \\
        \midrule
        \multirow{2}{*}{$0.3 < y < 0.5$} & \multirow{2}{*}{9-12} & $a/(z-c)^2$ & 4 & 6 \\
        & & $a/(z^2+bz)^c$ & 5 & 2  \\
        \midrule
        \multirow{2}{*}{$0.5 < y < 0.7$} & \multirow{2}{*}{5} 
        & $a/(1-b\cos(3z))^n$ & 6 & 6\\
        & & $a/(z^2 +bz)^n)$ & 1 & -\\
        & & $\exp(a(z-b(z-c)^2))/z$ & - & 1 \\
    \midrule
     Total & & & 38/38 & 38/38 \\
    \bottomrule
    \end{tabular}
    \label{tab:results_hpm}
\end{table}
\begin{figure*}[t]%
\centering
\includegraphics[width=8cm,height=5cm]{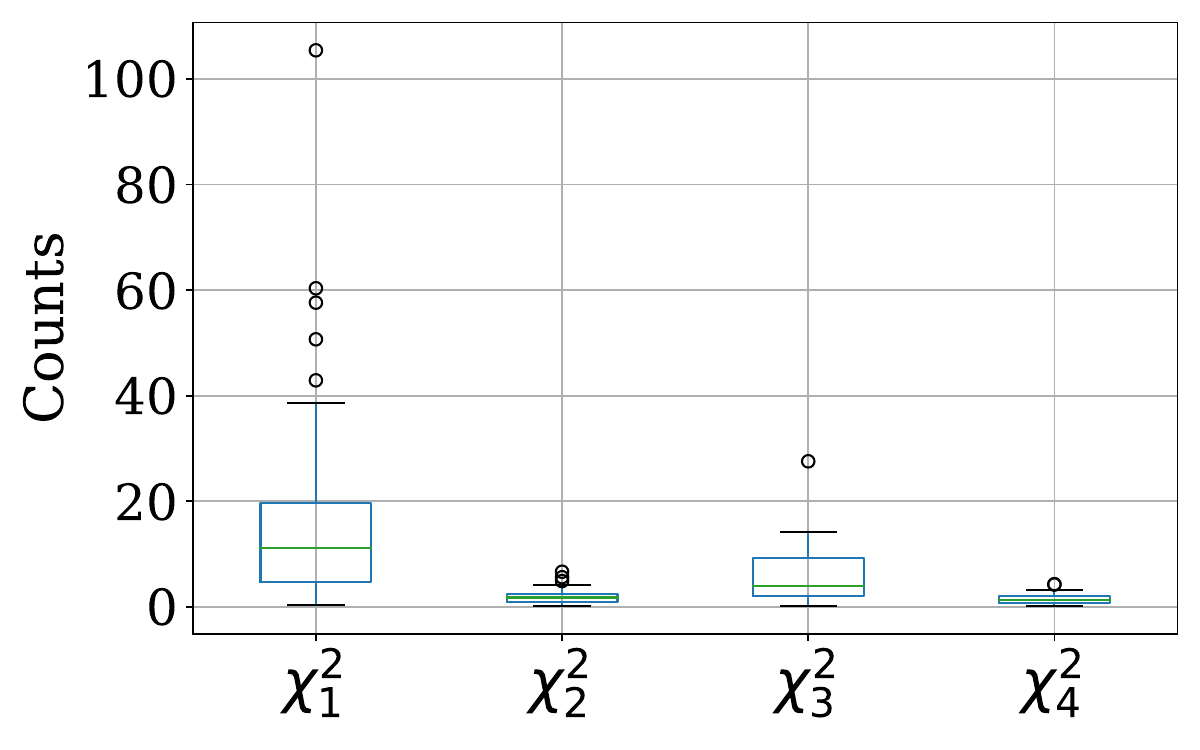}
\caption{Comparison of the fits of $h^{+}$ multiplicties~\cite{COMPASS:2016xvm} using the functions in Eq.~\ref{eq:srmodels} where $\chi^2_i$ denotes the $\chi^2/\mathrm{ndf}$ values obtained using $g_i(z)$. The box delimits the first and the third quartiles, whereas the middle line represents the median. The bottom and top lines represent, respectively, the minimum and maximum values in the $\chi^2/\mathrm{ndf}$ values. Markers show the outliers (values significantly smaller or larger than median values).}
\label{fig:compare_fits}
\end{figure*}
The function $f_2(z)$ outperforms $f_1(z)$ in describing the $z$ dependence of $h^{\pm}$ multiplicities, particularly at high $z$ ($z>0.5$). The best description of the data is obtained for $h^{-}$ in the range $x\in[0.02,0.03]$, with a loss of $5.8\times10^{-5}$, and numerical constants $a\approx-9.8$, $b=-7.7$, and $c\approx1$. This result is equivalent to the equation $f_2(z;x_3,y_3) = a(1-z)^{-1}\exp(-b z)$, which we will refer to as $f_3(z)$. The latter resembles the Lund FF (Eq.~\ref{eq:LundFF}) in the exponential factor and the component ($1-z$)$^{\alpha}$ with $\alpha=-1$. 
The multiplicity values in the corresponding bin, i.e., $0.2 < y< 0.3$ and $0.02 < x < 0.03$, are considered the ``training'' data in the context of this study, where a pre-trained transformer network is used. The multiplicity values in the remaining bins of the phase space thus represent the ``test'' data which plays a key role in ML-related studies. Test data allows for evaluating the overall performance of the learned model using ``unseen'' data, i.e., data points that were not part of the ``training'' set, and to check the generalizability of the SR model across the whole phase space (i.e., other ($x,y$) bins), and to other related measurements (e.g., $\pi$ and $K$ multiplicities, different energy scale and various HEP processes, etc.).

\begin{figure}[htp]%
\centering
\includegraphics[width=16.5cm,height=9cm]{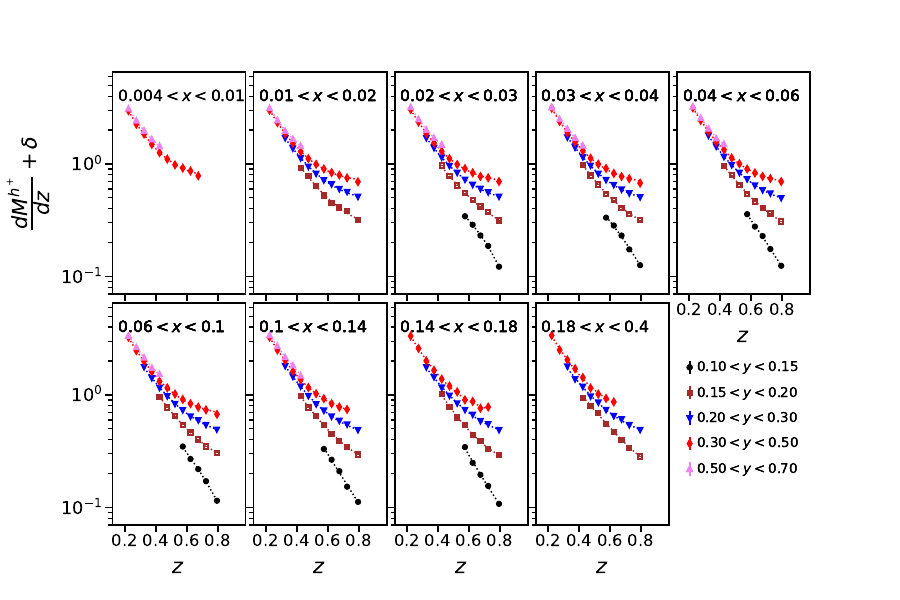} 
\caption{Comparison between experimental data~\cite{COMPASS:2016xvm} and fits performed using the SR model $g_4(z)$ (Eq.~\ref{eq:srmodels}) for positive hadron multiplicities, displayed as a function of $z$ in nine $x$ bins and five $y$ bins (staggered vertically by $\delta=0.3$ for clarity). Statistical uncertainties are considered in the fits, and $\chi^2/\text{ndf}$ values are summarized in Tab.~\ref{tab:chi2_pi_kp} (top-left).}
\label{fig:hpm_mul}
\end{figure}

The traditional ML approach achieves this goal by using $f_3(z) \equiv f_2(z;x_3,y_3)$ with the same numerical values of the constants $a$, $b$, and $c$ to make predictions on ``test'' data. For physics data, however, numerical constants in the models are more likely physical constants, which do not necessarily have the same values across the whole phase space and may exhibit weak or strong dependence upon kinematic variables. In fact, an important aspect of such multi-dimensional experimental measurements in physics is to investigate the dependence of physical constants upon the kinematic variables of interest in the measurement.
Therefore, to evaluate the performance of the learned models by SR on`` test'' data, we performed fits of $M^{h}(z)$ in individual kinematic bins. We consider the most frequently learned functional forms ($g_1\equiv f_1, g_2\equiv f_2$) and the the top-learned function $(g_3\equiv f_3)$. In addition, we consider a general form of $f_3$ by taking the power exponent in the term $(1-z)$ as a free fit parameter, referred to as $g_4$. This choice is mainly driven by the existence of a power exponent ``2'' in the learned function $f_1$. Thus, merging $f_1$ and $f_3$ into a general form requires the freeing of the exponent parameter. The list of fit functions thus includes:
\begin{equation}\label{eq:srmodels}
\begin{split}
    & g_1(z) = a\exp(-bz)/z^2 \hspace{2.1cm} g_3(z) = a\exp(-bz)/(1-z)\\
    & g_2(z) = a\exp(-bz)/(c-z) \hspace{1.25cm} g_4(z) = a(1-z)^{c}\exp(-bz)
\end{split}
\end{equation}
The best fits of $M^{h^{\pm}}$ are obtained using $g_2(z)$ and $g_4(z)$, both having an extra parameter $c$ and an overall better description over the phase space obtained using $g_4(z)$.
This is illustrated in Fig.~\ref{fig:compare_fits}, where a comparison of the range of $\chi^2/\mathrm{ndf}$ values obtained using different fit functions (Eq.~\ref{eq:srmodels}) is shown for $h^{+}$. 
We thus use $g_4(z)$ to fit multiplicities of charged hadrons, pions, and kaons in individual ($x,y$) bins to check the generalizability of the learned model i) within the same $h^{\pm}$ dataset (out-of-distribution) and ii) across hadron species, i.e., $\pi^{\pm}$ and $K^{\pm}$ (out-of-sample).
Table
~\ref{tab:chi2_pi_kp} summarizes the $\chi^2/\mathrm{ndf}$ values of $M^{h}$ fits performed in individual bins using $g_4(z)$, for ${h^{+}}$ and ${h^{-}}$, $\pi^{+}$, and $K^{+}$, providing strong evidence of the reliability of the learned function.
Figure~\ref{fig:hpm_mul} compares the $z$ dependence of measured multiplicities and predictions obtained using $g_4(z)$ in ($x,y$) bins for $h^{+}$. The fits, displayed by the dashed curves, significantly describe remarkably well the $z$ dependence of $M^{h^{+}}$ across all ($x,y$) bins. The same observation is obtained for positively~(Fig.~\ref{fig:pipm_kpm_mul}) and negatively charged pions, and kaons. 

\begin{table}
    \centering
    \caption{$\chi2$/ndf-values of the fits to $M^{h^{+}}$ and $M^{h^{-}}$ using $g_4(z)$ (cf. Eq.~\ref{eq:srmodels}).}
    \begin{tabular}{p{1.5cm}|p{1cm}p{1cm}p{1cm}p{1cm}p{1cm}|p{1cm}p{1cm}p{1cm}p{1cm}p{1cm}}
    \toprule
        & \multicolumn{10}{c}{$\chi^2/\text{ndf}$} \\ \hline
        & \multicolumn{5}{c|}{$h^{+}$} & \multicolumn{5}{c}{$h^{-}$} \\ \hline
        $x$ $\downarrow$ $y$ $\rightarrow$& $y_1$ & $y_2$ & $y_3$ & $y_4$ & $y_5$ & $y_1$ & $y_2$ & $y_3$ & $y_4$ & $y_5$ \\
        \hline
        $x_1$	&	&	&	&1.914	&0.451	& 
        &	&	&1.498	&4.820	\\
        $x_2$	&	&1.732	&3.723	&3.078	&1.493	&
        &0.703	&8.117	&3.894	&0.998	\\
        $x_3$	&6.646	&1.846	&0.823	&2.037	&0.529	
        &2.814	&1.400	&2.356	&1.396	&0.261	\\
        $x_4$	&5.607	&0.513	&0.410	&2.227	&0.236	
        &1.966	&1.418	&0.922	&4.377	&4.879	\\
        $x_5$	&1.850	&0.780	&1.141	&1.224	&4.190
        &2.613	&0.991	&1.612	&1.355	&2.714	\\
        $x_6$	&4.117	&0.958	&2.241	&2.715	&2.326	
        &0.123	&1.131	&2.073	&1.456	&3.684	\\
        $x_7$	&2.414	&0.661	&1.499	&0.333	&0.900	
        &5.641	&1.016	&2.329	&2.208	&0.115	\\
        $x_8$	&0.656	&1.050	&1.686	&1.807	&	
        &2.447	&1.268	&1.360	&1.751	&	\\
        $x_9$	&	&4.879	&2.391	&2.187	&	
        &	&0.523	&1.394	&1.144	&	\\
    \bottomrule
        & \multicolumn{5}{c|}{$\pi^{+}$} & \multicolumn{5}{c}{$K^{+}$} \\ \hline
        $y$-bin $\rightarrow$& $y_1$ & $y_2$ & $y_3$ & $y_4$ & $y_5$ & $y_1$ & $y_2$ & $y_3$ & $y_4$ & $y_5$ \\
        \hline
        $x_1$	&	&	&	&2.438	&0.930	&	&	&	&0.400	&1.179	\\
        $x_2$	&	&1.521	&5.551	&2.947	&1.365	&
        &1.262	&0.807	&2.459	&0.942	\\
        $x_3$	&0.479	&2.470	&2.776	&2.020	&0.101
        &0.812	&1.245	&0.854	&2.961	&4.732	\\
        $x_4$	&2.947	&1.157	&0.828	&1.844	&1.094
        &1.463	&0.979	&1.317	&1.557	&1.928	\\
        $x_5$	&1.990	&0.518	&1.373	&1.688	&0.147
        &2.717	&0.949	&1.239	&2.094	&0.805	\\
        $x_6$	&1.498	&1.206	&2.718	&2.735	&1.194
        &0.681	&1.147	&1.692	&0.622	&0.584	\\
        $x_7$	&1.825	&0.977	&3.129	&0.810	&0.874
        &0.262	&1.340	&0.957	&1.893	&1.367	\\
        $x_8$	&2.975	&1.984	&1.652	&2.353	&
        &1.892	&0.783	&0.806	&1.536	&	\\
        $x_9$	&	     &0.485	&2.945	&1.433	& 
        &	&2.291	&1.133	&0.811	&	\\
    \bottomrule
    \end{tabular}
    \label{tab:chi2_pi_kp}
\end{table}

\begin{figure}[htp]%
\centering
\includegraphics[width=16cm,height=12cm]{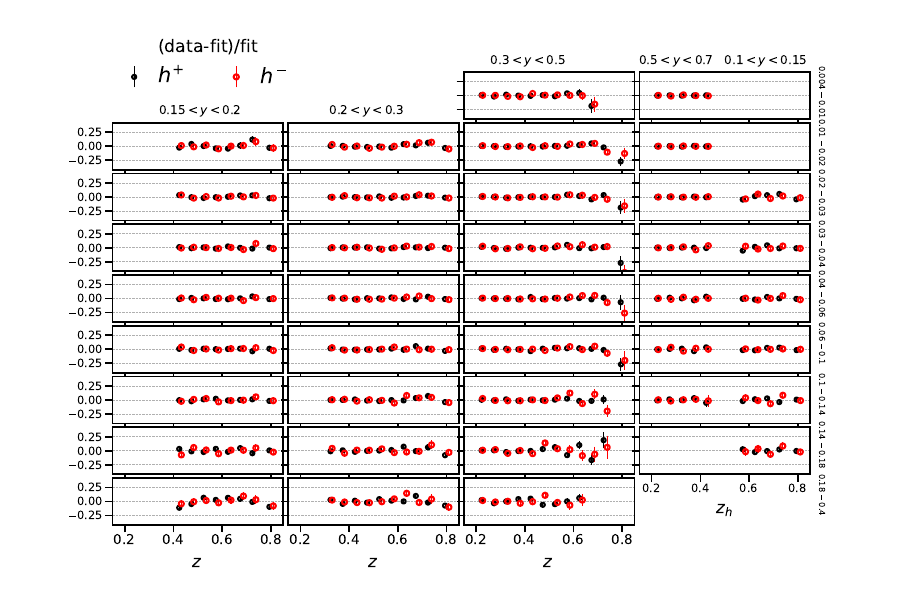} 
\caption{"(Data-fit)/fit" for the fit to the $h^{\pm}$ SIDIS multiplicities from~\cite{COMPASS:2016xvm} using $g_4(z)$.}
\label{fig:hpm_ratio}
\end{figure}

\begin{figure*}
    \centering
    \begin{subfigure}
        \centering
        \includegraphics[height=1.55in]{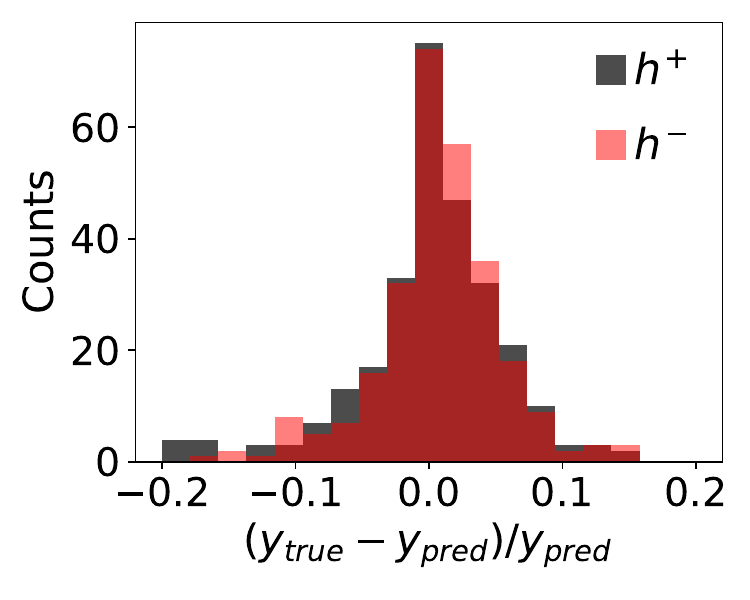}
    \end{subfigure}%
    ~ 
    \begin{subfigure}
        \centering
        \includegraphics[height=1.55in]{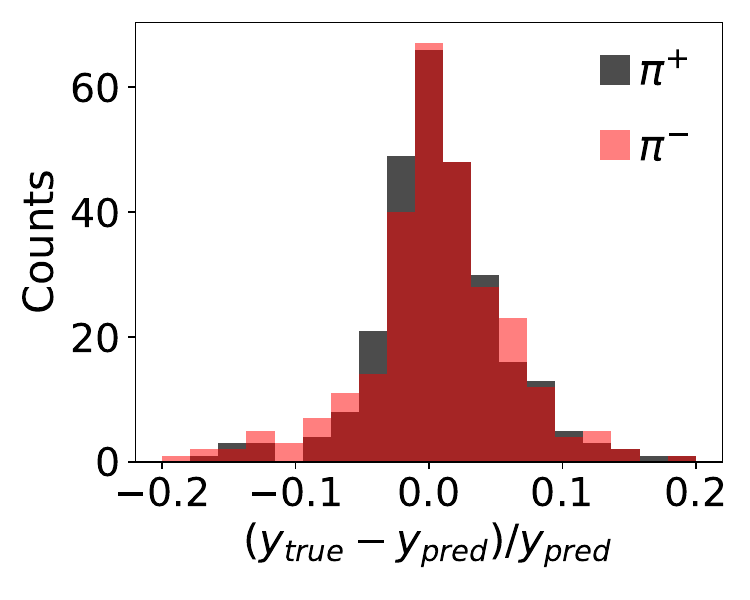}
    \end{subfigure}%
    ~ 
    \begin{subfigure}
        \centering
        \includegraphics[height=1.55in]{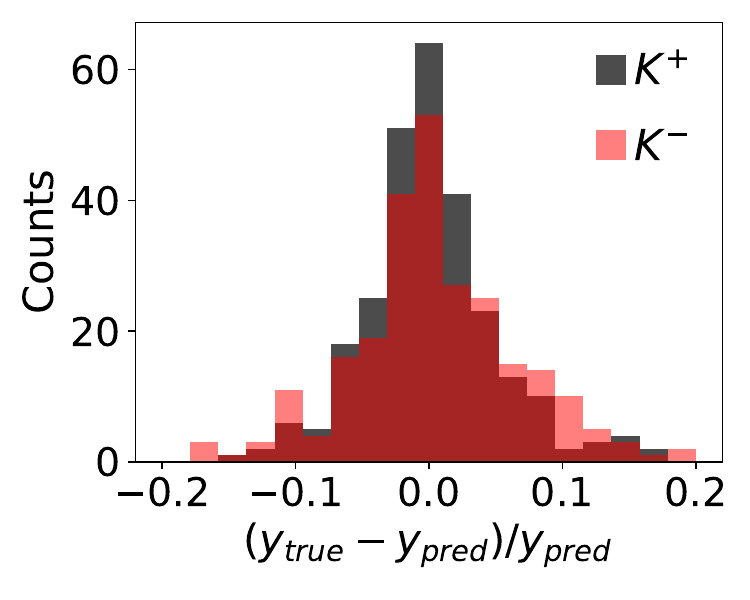}
    \end{subfigure}
    \begin{subfigure}
        \centering
        \includegraphics[height=1.55in]{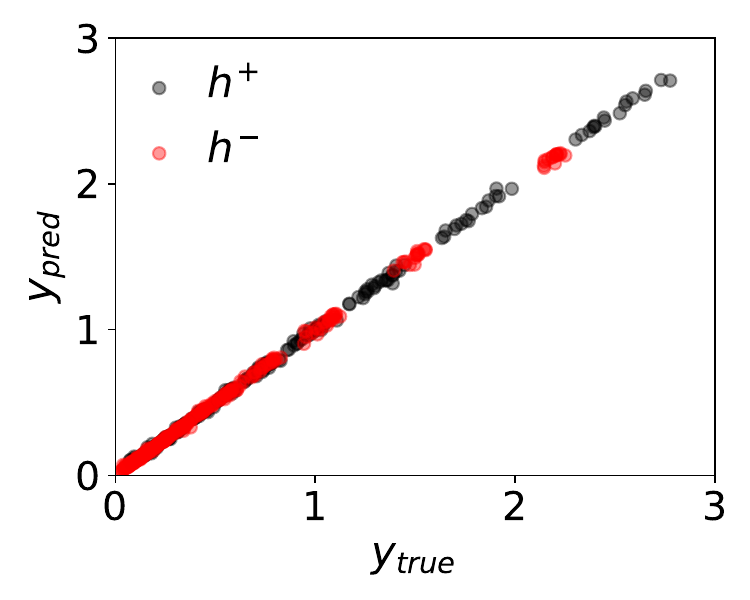}
    \end{subfigure}%
    ~ 
    \begin{subfigure}
        \centering
        \includegraphics[height=1.55in]{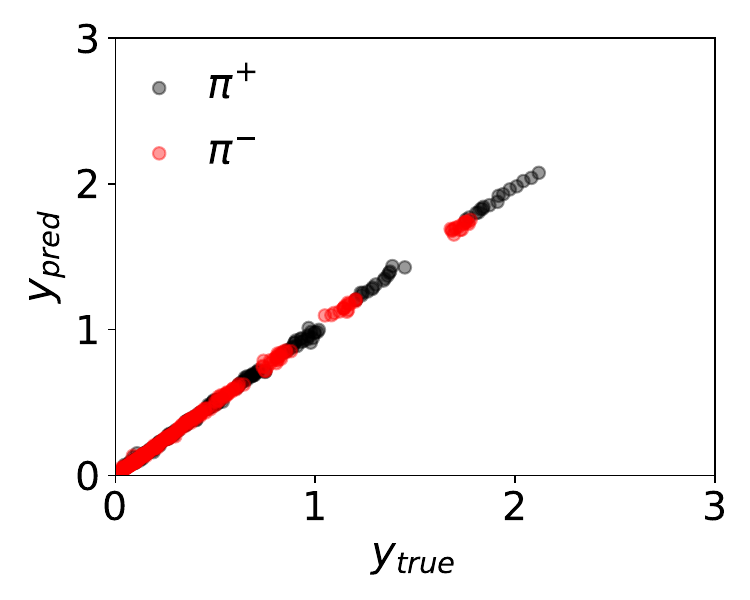}
    \end{subfigure}%
    ~ 
    \begin{subfigure}
        \centering
        \includegraphics[height=1.55in]{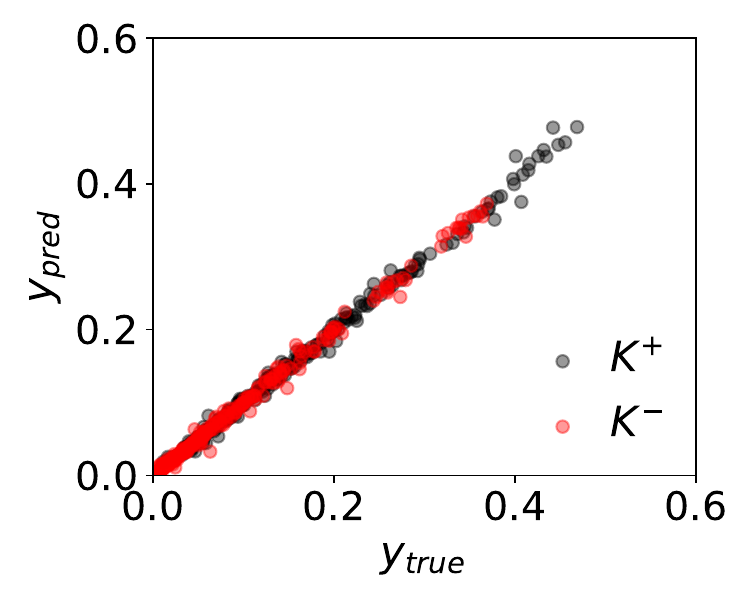}
    \end{subfigure}
    \caption{Fits' performance evaluated on the test set in terms of the distribution of relative errors for $h^{+}$, $\pi^{+}$, and $K^{+}$. For each case, data includes all measured points in ($x,~y,~z$) bins.}
    \label{fig:fit_quality}
\end{figure*}

Figure~\ref{fig:hpm_ratio} presents a “(data-fit)/ fit” comparison as a function of $z$ in individual ($x, y$) bins for $h^{\pm}$. The quality of the fits is remarkably good and comparable between positively and negatively charged hadrons. 
The ratios are \st{all} compatible with zero, and no systematic dependence upon $z$ is observed. The multiplicities in the highest $z$ bin are not correctly reproduced by the fits in the fourth $y$ bin, i.e., $0.3<y<0.5$; however, this observation does not hold for other $y$ bins. The same conclusion is found for $\pi^{\pm}$ and $K^{\pm}$ as shown by the ratios in Fig.~\ref{fig:pipm_kpm_ratio}. The overall distributions of the relative errors of the fits is shown in Fig.~\ref{fig:fit_quality} (upper panel) for $h^{\pm}$ (left), $\pi^{\pm}$ (center), and $K^{\pm}$ (right). A narrower distribution is obtained for hadrons compared to pions and kaons, which could be explained by the statistical significance of the unidentified hadron sample. This is also illustrated in Fig.~\ref{fig:fit_quality} (lower panel) which presents the “true vs. prediction” scatter plots for $h^{\pm}$, $\pi^{\pm}$, and $K^{\pm}$. 

\textbf{Analysis of charged pions and kaons.} We perform the SR analysis on differential multiplicities of charged pions and kaons. The majority of the learned equations, in these cases, include a trigonometric function, mostly having the functional form $a\cos(z)^2/z^2$ and $a\cos(z)^2/(z-b)$, and do not describe either the multiplicities or their $z$ dependence. They are thus not reported here. This observation questions the inner mechanism through which SR learns a model, especially that hadron and pion multiplicities are very similar in terms of $z$ dependence, but also demonstrates the importance of and the need for a multidimensional measurement in such applications, where a functional form could be learned in only a few intervals among all but generalizes well to other intervals of the phase space and to other related measurements.\\
As previously mentioned, the main goal of this analysis is to learn a functional form of FFs directly from data, which could then be used in global QCD fits where measurements at different energy scales from various high-energy processes are considered to learn FFs. 
The top learned SR model $g_4(z)$ is expected to express the underlying mechanism in data and thus represent FFs. This is supported by the striking similarity between the Lund FF (Eq.~\ref{eq:LundFF}) and $g_4(z)$, especially in the $(1-z)^c$ component, which enforces the FF constraint $D_i^h(z=1, Q) = 0$. Finally, the top learned function $g_4(z)$ could be regarded as a potential candidate to serve as a FF's parameterization in global QCD fits.

\vspace{-.4cm}
\subsection{{Learning univariate function from fragmentation function values}}\label{sec:1dff}
\vspace{-.25cm}

A direct point-by-point extraction of quark-to-hadron FFs is possible at leading-order (LO) in perturbative QCD (pQCD) using multiplicities of identified charged hadron (i.e., $\pi$ and $K$) measured in SIDIS, without the need for a pre-assumed functional form for FFs. A hadron multiplicity is defined by the ratio of the SIDIS and DIS cross sections which are expressed at LO in pQCD in terms of the parton distribution functions (PDFs) and FFs and read as follows:
\begin{equation}\label{eq:xsections}
\begin{split}
& \frac{\dd^2\sigma^{\mathrm{DIS}}}{\dd x\dd Q^2} = C(x, Q^2) \sum_q e_q^2 f_q(x, Q^2), \\
& \frac{\dd^3\sigma^h}{\dd x\dd Q^2\dd z} = C(x, Q^2) \sum_q e_q^2 f_q(x, Q^2)D_q^h(z, Q^2).
\end{split}
\end{equation}
Here, $C(x,Q^2)$ is a kinematic factor, $f_q(x, Q^2)$ is the PDF for the quark flavour $q$, and $D_q^h(z, Q^2)$ the FF od the quark $q$ into a hadron $h$. The kinematic factor $C(x, Q^2)$ is eliminated in the ratio $\sigma^{\text{SIDIS}}/\sigma^{\text{DIS}}$, and both the numerator and denominator reduce to linear combinations of FFs where the PDFs are evaluyated at the corresponding $x$ and $Q^2$ values. The $\pi^{\pm }$ multiplicity, for example, can be written as:
\begin{equation}
    M^{\pi^{+}(\pi^-)} = \frac{\sum_q e_q^2 q(x,Q^2)D_q^{\pi^+(\pi^-)}(z,Q^2)}{\sum_q e_q^2 q(x,Q^2)}
\end{equation}
The fragmentation of a quark of a given flavour into a hadron is called favoured if the quark flavour corresponds to a valence quark in the hadron. Otherwise the fragmentation is called unfavoured.  For the pion case, and according to isospin and charge symmetry, three independent quark-to-pion FFs remain at LO: i) favoured FF $D_{\text{fav}}^{\pi}$, ii) unfavoured FF $D_{\text{unf}}^{\pi}$, and iii) strange FF $D_{\text{str}}^{\pi}$, such that:
\begin{equation}
\begin{split}
    D_{\mathrm{fav}}^{\pi} &= D_{u}^{\pi^+} = D_{\bar{d}}^{\pi^-} = D_{d}^{\pi^+} = D_{\bar{u}}^{\pi^-} \qquad D_{s}^{\pi^{\pm}} = D_{\bar{s}}^{\pi^{\pm}}.\\
    D_{\mathrm{unf}}^{\pi} &= D_{d}^{\pi^+} = D_{\bar{u}}^{\pi^-} = D_{u}^{\pi^+} = D_{d}^{\pi^-}
\end{split}
\end{equation}

In each $(x, y, z)$ bin, the problem reduces to solve a system of two linear equations for $M^{\pi^+}$ and $M^{\pi^-}$ for the values of 
$D_{\text{i}}^{\pi}((z), \langle Q^2 \rangle)$, where $i\in\{\text{fav, unf, str}\}$, and $\langle Q^2\rangle$ denotes the mean values of $Q^2$. No functional form has to be assumed, the DGLAP evolution for FFs is not required as in global QCD fits, and MSTW08 at LO~\cite{Martin_2009} is selected for the PDFs. The extracted set of FFs ($D_{\text{fav}}, D_{\text{unf}}, D_{\text{str}}$) is then passed into the SR algorithm~\cite{nesymres} to learn the underlying equations. 
This exercise not only allows for checking the consistency of the functional form of the underlying mechanism that is inferred from multiplicities but also to verify the validity of an LO extraction of FFs. Such a test was never performed, although this is not the main goal of this study.  The top-performing functions are:
\begin{equation}
\begin{split}\label{eq:all_ff_fcts}
    & h_1(z) = a/(z-b)^2 \\
    & h_2(z) = a\exp(-bz)/z^2
\end{split}
\end{equation}
These two functions could be regarded as equivalent forms of the following function, for particular values of the parameters $b$ and $c$:
\begin{equation}\label{eq:ff_fit}
    h(z) = a\exp(-bz)/(z-c)^2
\end{equation}

We fit the extracted FFs using the functions in Eqs.~\ref{eq:all_ff_fcts} and~\ref{eq:ff_fit} in individual ($x,y$) bins. The function $h(z)$ remarkably outperforms $h_1(z)$ and $h_2(z)$ in terms of $\chi^2$ and provides a better description of the extracted FFs. This reflects that SR could capture patterns in some kinematic bins and other patterns in other bins such that the combination of the learned equations provides an overall better description in the fully covered phase space.
The similarity between $h(z)$ and $g_4(z)$ is remarkable, and the difference in the power index could be explained by stating that both share the structure $a\exp(-bz)(c-z)^{d}$ with different values of $c$ and $d$. However, $h(z)$ does not fulfill the FF constraint at $z=1$.

The observables considered in this study are  multiplicities of charged hadrons that measured in SIDIS, which can be expressed as a linear combination of FFs at LO in pQCD. The models inferred by SR are therefore functional forms of FFs, and this is reflected in the consistency of the results obtained using multiplicity data and FFs extracted distributions. This suggests that the learned function would generalize to other types of high-energy physics data, such as electron-positon and proton-proton collisions, since the SR output is a model of hadron FFs and not hadron multiplicity, and given the universality of FFs.

\vspace{-.4cm}
\subsection{{Learning bivariate function from multiplicity values}}\label{sec:2d}
\vspace{-.25cm}

\begin{figure}[htp]%
\centering
\includegraphics[width=16cm,height=11.5cm]{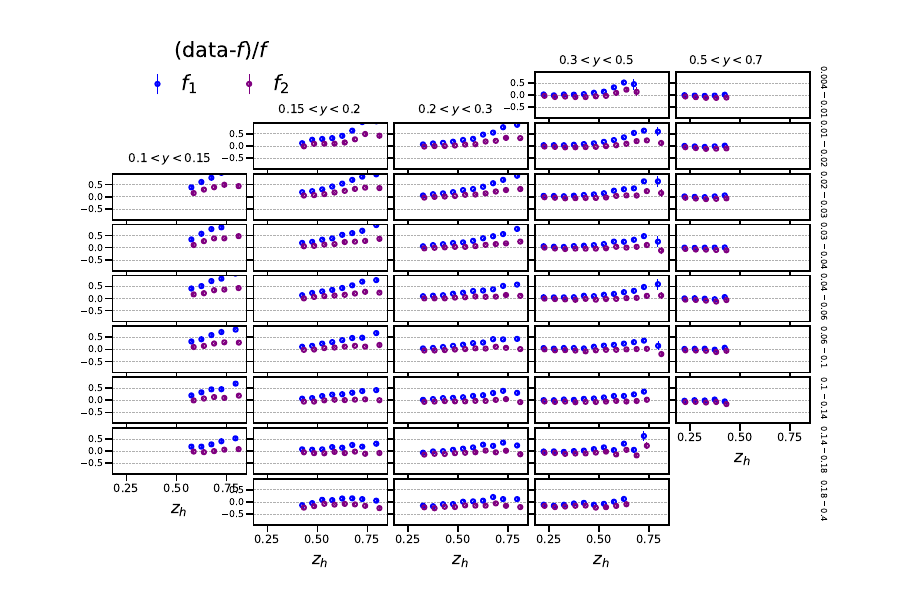} 
\caption{Comparison between $h^{+}$ multiplicity values~\cite{COMPASS:2016xvm} and those predicted using $f_1=f(z,x)$ (cf. Eq.~\ref{eq:2d}) and $f_2=(1-z)^{-0.2}f(z,x)$ in terms of (Data-$f$)/$f$, using the numerical values of the parameters that were learned by SR. No fit has been performed to produce this figure.}
\label{fig:f12}
\end{figure}

\begin{figure}[]%
\centering
\includegraphics[width=16cm,height=11.5cm]{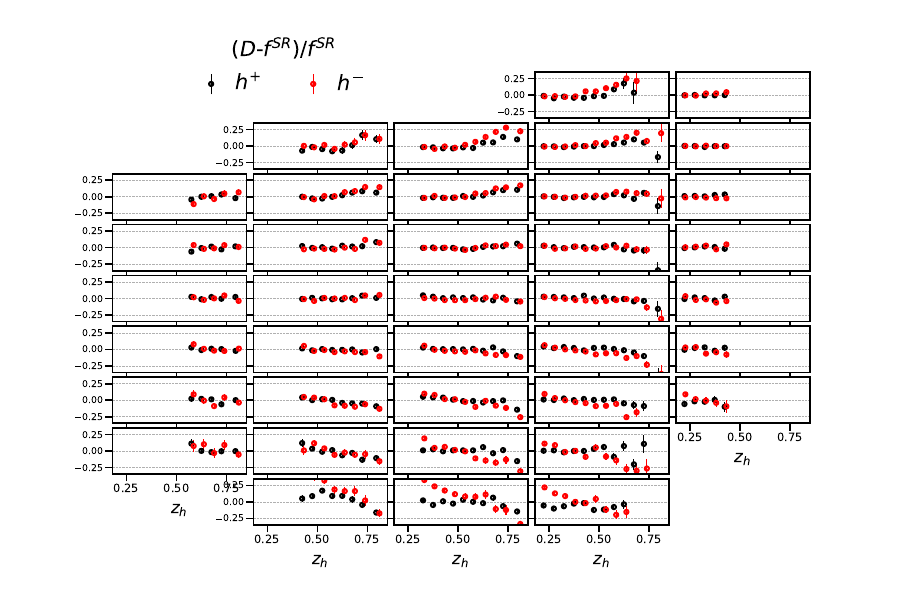} 
\caption{Comparison between multiplicity values~\cite{COMPASS:2016xvm} and those predicted using $f^{\prime}(z,x)$ in terms of (Data-$f$)/$f$ }
\label{fig:comp_data_2dfits}
\end{figure}

We apply SR (\textit{NeSymRes}) to two-dimensional multiplicities of charged hadrons $M^{h^{+}/h^{-}}(z,x)$ in five $y$ bins, resulting in five candidate equations for each of $h^{+}$ and $h^{-}$. The equations that depend on both variables are learned in the largest $y$ range, i.e., $0.5<y<0.7$, for both $h^{+}$ and $h^{-}$ with loss values of $7.3\times 10^{-4}$ and $1.2\times 10^{-3}$ respectively. This observation is intriguing because only five data points are measured in this $y$ range. They are:
\begin{equation}
   \begin{split}\label{eq:fxz}
        f^{h^{+}}(z,x) = \exp\left(-\alpha z + 2.26(1 + \beta x)^2\right) \quad \alpha\approx6.4,~\beta\approx 0.27 \\
        f^{h^{-}}(z,x) = \exp\left(-\alpha z + 2.39(1 - \beta x)^2\right) \quad \alpha\approx7.2,~\beta\approx 0.07
   \end{split} 
\end{equation}
%
The learned functions (Eq.~\ref{eq:fxz}) only differ by the sign of the $\beta$ parameter. The function associated with the lowest error is selected and is equivalent to the following:
\begin{equation}\label{eq:2d}
    f(z,x) = \exp(-6.4 z)\times \exp\left(2.3(1+0.27 x)^2\right)
\end{equation}
A first observation is a factorization in the dependence of $f(z,x)$ (Eq.~\ref{eq:2d}) upon $z$ and $x$ through the exponential, providing the first experimental evidence of the factorization theorem that is usually assumed in phenomenological studies of FFs. The next step is to check the validity of the learned model for ``test'' data (i.e., $0.15<y<0.5$). 
Figure~\ref{fig:f12} compares $h^{+}$ multiplicity values to those predicted using $f_1\equiv f(z,x)$ (cf. Eq.~\ref{eq:2d}) by taking the average values of $z$ and $x$ in each kinematic bin.
The comparison shows that $f_1$ correctly produces the multiplicity values for most of the $x$ ranges, particularly at the largest $x$ (last row). However, the $z$ dependence of $M^{h}$ is not correctly predicted, particularly for large values of $z$.
This discrepancy can be attributed to the absence of the term $(1-z)^{c}$, which was originally present in $g_4(z)$ (Eq.~\ref{eq:srmodels}), and stems from the constrained $z$ range (i.e., $z<0.5$) over which $f(z,x)$ was derived within the bin $0.5<y<0.7$. By explicitly adding this term to $f(z,x)$, the $z$ dependence of the multiplicities improves in all $x$ ranges, as illustrated in Fig.~\ref{fig:f12} where $f_2=(1-z)^{-0.2}f(z,x)$.\\
The ultimate goal is to learn a function that correctly predicts the multiplicity values and their $z$ dependence without having to fit the parameters' numerical values for each kinematic bin but instead using solely the values of the kinematic variables $x$ and $z$. We thus search for the numerical values of the parameters that could describe the $z$ and $x$ dependencies of the multiplicities in all ($x,y$) bins by fitting the two-dimensional multiplicity values $M^{h^+/h^-}(z,x)$ in five $y$ ranges using the two-dimensional function:
\begin{equation}\label{eq:2d_function}
    f^{\prime}(z,x) = N(1-z)^{\gamma}\exp\left(-\alpha z + 2.3(1-\beta x)^2\right)
\end{equation}
The comparison between the multiplicity values and the fits using Eq.~\ref{eq:2d_function} is shown in Fig.~\ref{fig:comp_data_2dfits}. It is evident that the numerical values of the fit parameters ($\alpha, \beta, \gamma$) obtained in a given $y$ bin do not properly describe the data across the $x$ bins, reflecting a dependence of the parameters upon $x$ and $y$. 
The numerical values of the parameters are summarized in Tab.~\ref{tab:par2d}. They clearly exhibit a dependence upon $y$, preventing us from finding common values and allowing us to make predictions solely on the basis of $x$ and $z$ mean values per bin.
\begin{table}[t]
    \centering
    \begin{tabular}{c|cccccccc|}
    \toprule
    $y$ bin & $\chi^2/$ndf & $N\pm\sigma_{N}$ & $\gamma\pm\sigma_{\gamma}$ & $\alpha\pm\sigma_{\alpha}$ & $\beta\pm\sigma_{\beta}$ \\
    \midrule
    1 & 0.85 & 0.327 $\pm$ 0.066 &  0.547 $\pm$ 0.172 &  2.980 $\pm$ 0.591 &  -0.249 $\pm$ 0.030 \\ 
    2 & 3.10 & 0.691 $\pm$ 0.061 &  0.094 $\pm$ 0.121 &  5.034 $\pm$ 0.336 &  -0.080 $\pm$ 0.022 \\  
    3 & 4.85 & 0.842 $\pm$ 0.030 &  -0.191 $\pm$ 0.076 &  5.950 $\pm$ 0.178 &  0.027 $\pm$ 0.017 \\  
    4 & 3.82 & 1.040 $\pm$ 0.017 &  -1.029 $\pm$ 0.092 &  7.755 $\pm$ 0.161 &  0.177 $\pm$ 0.013 \\
    5 & 1.07 & 1.193 $\pm$ 0.044 &  -2.931 $\pm$ 0.511 &  10.719 $\pm$ 0.735 &  0.336 $\pm$ 0.021 \\
    \bottomrule
    \end{tabular}
    \caption{Values of the parameters ($N, \gamma, \alpha, \beta$) from fits to two-dimensional $M^{h}(z,x)$ using $f^{\prime}(z,x)$ (cf. Eq.~\ref{eq:2d_function}).}
    \label{tab:par2d}
\end{table}

This exercise is repeated for pion multiplicities, i.e., $M^{\pi^{+}/\pi^{-}}(z,x)$. A two-dimensional function is learned for $\pi^{-}$ in the last $y$ range with a loss of $0.22$ (in comparison to $\approx 10^{-3}$ for $h^{+}/h^{-}$) and is $f^{\pi^{-}}(z,x)=0.9/(1 + 0.038*\cos(z/x))^{1.14}$. Learning $f^{\pi^{-}}(z,x)$ in the last $y$ range is consistent with the previous finding. However, the inferred function is not considered, given that there is no periodic behavior expected in such data.

\vspace{-.4cm}
\section{Key Learnings}
\vspace{-.2cm}

Our study reveals important insights about applying SR to complex physics data. The analysis progressed through different levels of dimensional reduction, yielding several key findings. The full three-dimensional dataset ($\mathrm{x}=\{z,x,y\}$) proved too complex for successful SR. However, reducing to two-dimensional subsets ($\mathrm{x}=\{z,x\}$) yielded partial success, allowing us to identify some meaningful functional relationships. Further reduction to one-dimensional analysis ($\mathrm{x}=\{z\}$), specifically focusing on the relationship between $z$ and $M^{h}$, produced a robust z-dependent function that successfully fit the data across all kinematic bins. Notably, this function maintained its predictive power even in bins where it wasn't directly trained.

An interesting observation emerged: certain mathematical terms discovered in the simpler $\{z,M^{h}\}$ analysis weren't recovered when analyzing higher-dimensional datasets. This suggests a hierarchical learning process where simpler, fundamental relationships may be obscured in higher-dimensional analyses, and vice versa. These findings highlight the value of incorporating domain knowledge when determining the optimal level of dimensional reduction for symbolic regression analysis.

\vspace{-.4cm}
\section{Conclusion}
\vspace{-.2cm}

This paper presents the first study where a functional form of FFs is inferred directly from data, using symbolic regression. FFs have always been determined in global QCD fits using pre-defined functional forms. We consider two alternatives to learn FFs' functional form, using charged hadron multiplicities and $z$ dependent distributions of FFs extracted from charged pion multiplicities at LO in pQCD. The resulting function is:

\begin{equation}
    f^{\text{SR}} = a(1-z)^{c}\exp(-bz)
\end{equation}

The function $f^{\text{SR}}$ resembles the Lund symmetric fragmentation function (Eq.~\ref{eq:LundFF}) but distinct. The learned function is fit to data and found to describe them very well for all hadron species and over the whole phase space covered in the measurement. The learned function could be used in the next versions of global QCD fits, which comprise data from different high-energy physics experiments. This would be a departure from traditional methodology where in such case, both the model and its parameters would originate from data.
Finally, this result show a promise in using symbolic regression to learn mathematical equations direclty from experimental data, and could be applied to many sub-areas in high-energy physics and to other physics disciplines.

\appendix
\section{Additional Table and Figures}

\begin{table}[htp]
    \centering
    \begin{tabular}{cccccccccccccc}
    \toprule
        \multicolumn{14}{c}{Bin limits} \\
        \midrule
         $y$ & 0.1 & 0.15 & 0.2 & 0.3 & 0.5 & 0.7 & & & & & & &\\
         $x$ & 0.004 & 0.01 & 0.02 & 0.03 & 0.04 & 0.06 & 0.1 & 0.14 & 0.18 & 0.4 & & & \\
         $z$ & 0.2 & 0.25 & 0.3 & 0.35 & 0.4 & 0.45 & 0.5 & 0.55 & 0.6 & 0.65 & 0.7 & 0.75 & 0.85 \\
     \bottomrule
    \end{tabular}
    \caption{Bin limits for the kinematic variables $x$, $y$ and $z$ from~\cite{COMPASS:2016xvm}.}
    \label{tab:kinematics}
\end{table}

\begin{figure}[htp]%
\centering
    \begin{subfigure}
        \centering
        \includegraphics[width=15cm,height=8.5cm]{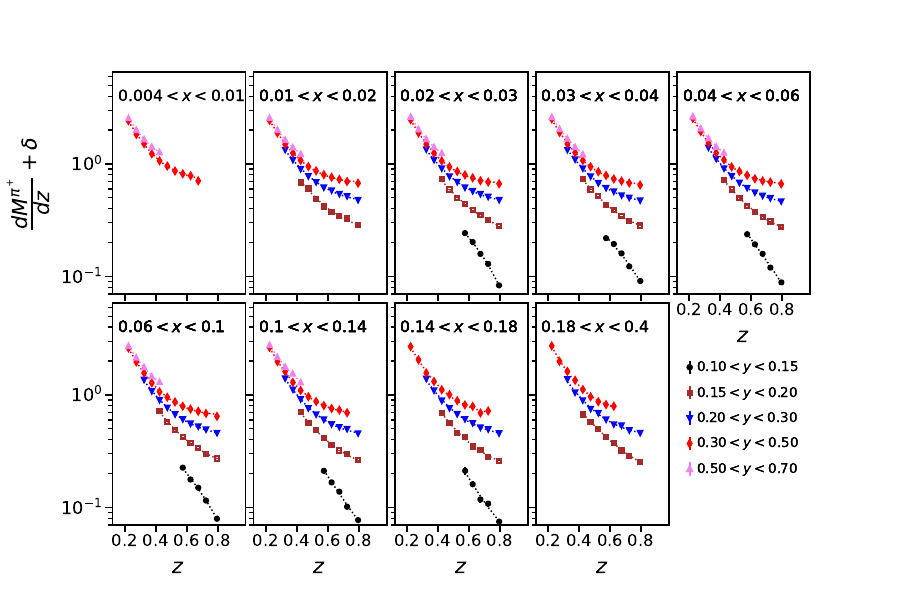} 
    \end{subfigure}
    \hfill
    \begin{subfigure}
        \centering
        \includegraphics[width=15cm,height=8.5cm]{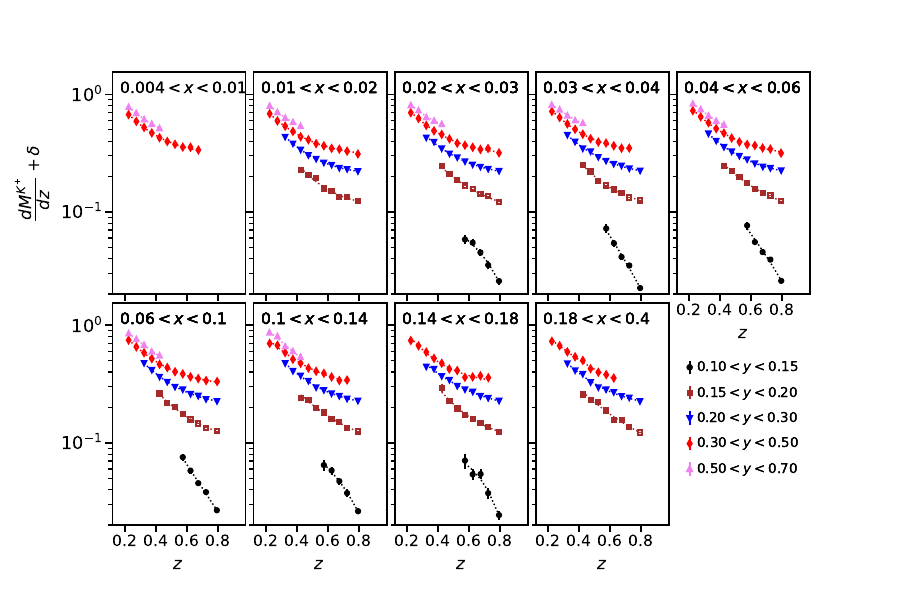}
    \end{subfigure}
\caption{Comparison between experimental data~\cite{COMPASS:2016xvm} and fits performed using the SR model $g_4(z)$ (Eq.~\ref{eq:srmodels}) for positive pion~\cite{COMPASS:2016xvm} and kaon~\cite{COMPASS:2016crr} multiplicities, displayed as a function of $z$ in nine $x$ bins and five $y$ bins (staggered vertically by $\delta=0.3$ for clarity). Statistical uncertainties are considered in the fits, and $\chi^2/\text{ndf}$ values are summarized in Tab.~\ref{tab:chi2_pi_kp} (bottom).}
\label{fig:pipm_kpm_mul}
\end{figure}
\begin{figure}[htp]%
\centering
    \begin{subfigure}
        \centering
        \includegraphics[width=14cm,height=10cm]{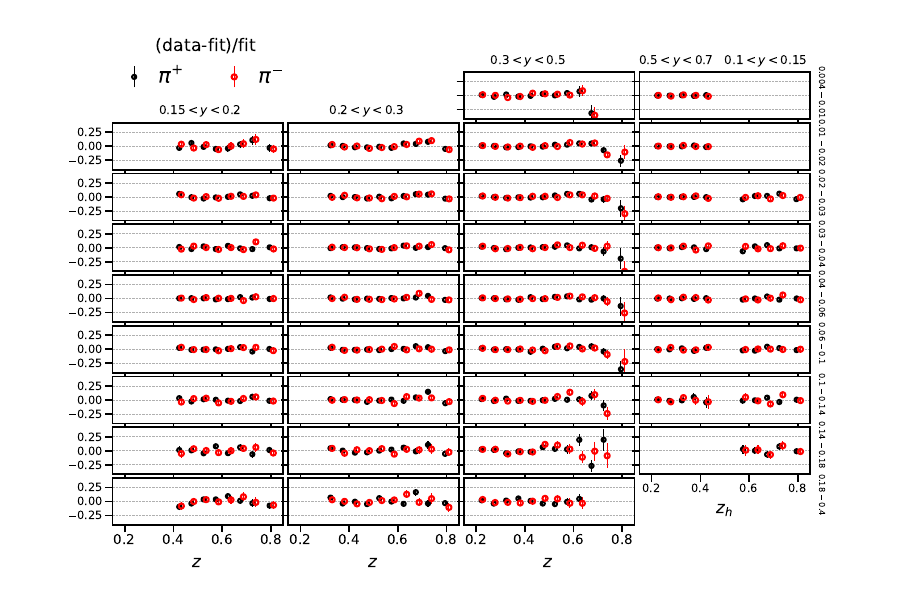} 
    \end{subfigure}
    \begin{subfigure}
        \centering
        \includegraphics[width=14cm,height=10cm]{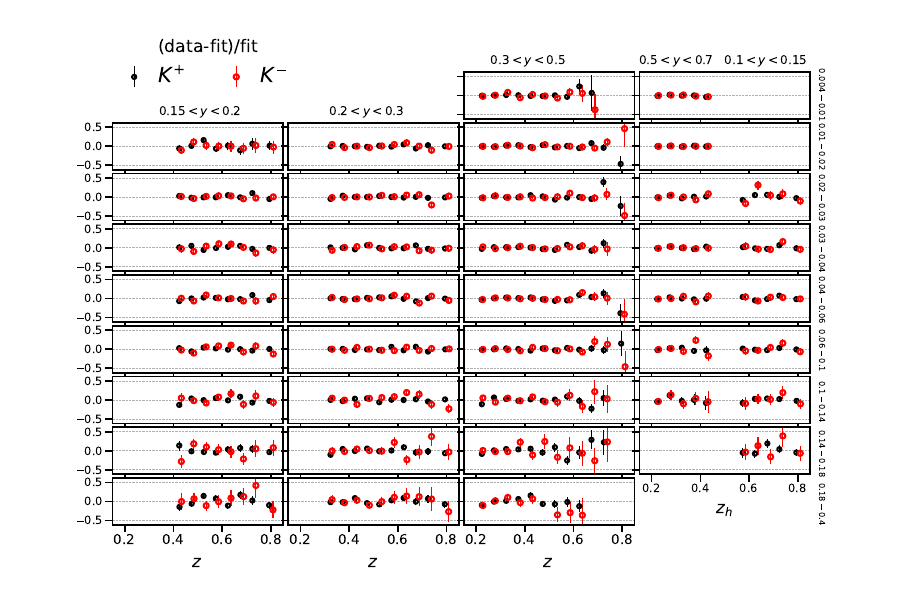}
    \end{subfigure}
\caption{"(Data-fit)/fit" for the fit to the $K^{\pm}$ SIDIS multiplicities from~\cite{COMPASS:2016crr} using $g_4(z)$.}
\label{fig:pipm_kpm_ratio}
\end{figure}


\bibliographystyle{vancouver}
\bibliography{sample}

\end{document}